\documentclass[9pt,journal,compsoc]{IEEEtran}

\ifCLASSOPTIONcompsoc
\usepackage[nocompress]{cite}
\else
\usepackage{cite}
\fi

\usepackage{mathptmx} 

\newcommand{\ignore}[1]{}
\usepackage{fancyhdr}
\usepackage[normalem]{ulem}
\usepackage{microtype}

\usepackage{booktabs} 
\usepackage{subfig}
\usepackage{paralist}
\usepackage{listings}
\usepackage{todonotes}
\usepackage{caption}
\usepackage{booktabs}
\usepackage{multirow}
\usepackage[hyphens]{url}
\usepackage{array}
\usepackage{amsmath}
\usepackage[ruled,vlined,linesnumbered]{algorithm2e}
\usepackage{flushend}

\usepackage[bookmarks=true,breaklinks=true,letterpaper=true,colorlinks,linkcolor=black,citecolor=blue,urlcolor=black]{hyperref}

\begin{document}
\title{Towards a Better Indicator for Cache Timing Channels}
	
\author{
		\vspace{-2mm}
		Fan~Yao,~\IEEEmembership{Member,~IEEE,}
		Hongyu~Fang,~\IEEEmembership{Student Member,~IEEE,}
		Milos~Doroslovacki,~\IEEEmembership{Member,~IEEE,}\\
		Guru~Venkataramani,~\IEEEmembership{Senior Member,~IEEE
		\vspace{-2mm}
	   }
}


\IEEEtitleabstractindextext{%
	\begin{abstract}
		Recent studies highlighting the vulnerability of computer architecture to information leakage attacks have been a cause of significant concern. Among the various classes of microarchitectural attacks, cache timing channels are especially worrisome since they have the potential to compromise users' private data at high bit rates. Prior works have demonstrated the use of cache miss patterns to detect these attacks. We find that cache miss traces can be easily spoofed and thus they may not be able to identify smarter adversaries. 
		In this work, we show that \emph{cache occupancy}, which records the number of cache blocks owned by a specific process, can be leveraged as a stronger indicator for the presence of cache timing channels. We observe that the modulation of cache access latency in timing channels can be recognized through analyzing pairwise cache occupancy patterns. Our experimental results show that cache occupancy patterns cannot be easily obfuscated even by advanced adversaries that successfully evade cache miss-based detection.
	\end{abstract}
	
	\begin{IEEEkeywords}
		Computer Security, Timing Channels, Cache Occupancy Analysis.
\end{IEEEkeywords}}

\maketitle

\IEEEdisplaynontitleabstractindextext

\IEEEpeerreviewmaketitle

\IEEEraisesectionheading{\section{Introduction}\label{sec:introduction}}

\IEEEPARstart{W}{ith} rapid growth in the use of computer systems for storing and accessing user data, protecting sensitive information and shielding them from malicious entities is an important task for computer architects. 
The recent attacks exploiting micro-architecture~\cite{spectremeltdown} have further stressed the need for hardware and information security to be considered as first-order design constraints in computer architecture.
Among the many forms of information leakage attacks, timing channels are particularly notorious for their stealthy exfiltration of sensitive information, leaving no physical evidence for forensic examination. These timing channels simply rely on the modulation of resource access timing that cannot be easily audited by software security monitors. 

Among various architectural units, caches are most exploited for timing channel attacks due to two major reasons: 1. CPU cache is one of the most commonly shared resources, and hence processes from different domains typically have access to it. 2. Caches are tightly coupled with processor pipelines and cannot be simply disabled for security reasons. 

Prior studies have studied cache-based timing channel attacks that manipulate accesses on various cache levels~\cite{liu2015last,yao_hpca18}.
CC-hunter~\cite{chen2014cc} detects covert timing channel by capturing cache conflict misses between two processes, and needs hardware to provide fine-grained information about mutual cache evictions. This may potentially involve increased hardware costs if such infrastructure is not already available. Recent works identify cache timing channel by analyzing the cache miss patterns using existing performance counters~\cite{chiappetta2016real,payer2016hexpads}. While these techniques may be more cost-effective without requiring additional hardware modifications, we demonstrate in this work that such cache miss-based detection could be evaded by sophisticated adversaries, and hence, may be circumvented.

In this paper, we propose the use of a new statistic--\emph{cache occupancy}--that can be utilized for improved detection of cache timing channels.
Cache occupancy records the number of cache blocks owned by a specific process in a certain cache during the observation period. We find that the trojan/victim and spy's modulation of cache access latencies generates unique patterns in cache occupancy traces between the two processes, that can be identified through correlation analysis. 
Meanwhile, unlike cache miss events that only offer a uni-dimensional view (i.e., always a non-negative number of events in the current epoch), cache occupancy profiles exhibit a gain-loss pattern depending on the cache blocks gained or lost in the current observation period. Essentially, cache occupancy patterns for two processes inherently reflect  mutual cache eviction behavior and can be more indicative of potential malicious behavior by a spy process (and/or trojan in case of covert timing channels). Our study demonstrates that cache occupancy is a much better indicator for cache timing channels.
Fortunately, recent commercial processors already started to support cache occupancy monitoring for application domains at runtime~\cite{intel-manual-2016}, making it possible to build practical and robust solutions for cache timing channel detection.

In summary, the major contributions of our article are:
\vspace{1mm}
\begin{itemize}
	\item We show that the adversaries' attempt to modulate cache access latencies using conflict misses generates distinct cache occupancy patterns during cache timing channels. By analyzing the cache occupancy profiles between two suspicious processes, cache timing channel may be inferred.
	
	\item We design and demonstrate a prototype for cache timing channel detection, and evaluate using real-world cache timing channel attacks. Our results show that, despite adversaries' effort to add intentional cache miss patterns and evade detection, our analysis using cache occupancy is still able to accurately capture trojan-spy communication. We highlight the potential of leveraging cache occupancy to build robust cache timing channel detection.
\end{itemize}

\section{Background and Threat Model}
\label{sec:background}

\subsection{Cache Timing Channel Attacks}
\label{sec:bg-attacks}

Timing channels typically involve two processes: trojan/victim and spy, where the spy learns of sensitive secrets from trojan/victim. Within many kinds of cache-based timing channels, Prime+Probe is the most commonly leveraged technique for cache access timing modulation~\cite{liu2015last}.
Typically, the spy deciphers secrets by measuring the cache access timing and finding out whether the accesses result in cache conflict misses or not. 
In general, there are two classes of communication protocols: single group-based attacks and multiple groups-based attacks~\cite{replayconfusion}.

\vspace{1mm}
\noindent\textbf{Single-group attacks}~\cite{whispers}: In this class of attacks, the trojan and spy sweep through the entire cache or partial of the cache. During the prime phase, the spy fills some cache sets. The trojan either accesses the cache sets to fill them with its own data, or remains idle and spy's contents are left intact. The spy probes these cache blocks and measures access latencies. Longer latency values indicate cache conflict misses, while shorter latencies indicate cache hits. Secret bits are deciphered based on cache latencies. 

\vspace{1mm}
\noindent\textbf{Multiple-group attacks}~\cite{liu2015last}: In these attacks, the trojan and spy exploit multiple groups (e.g., two groups) of cache sets to communicate the bits. Initially, the spy primes both groups of cache sets by filling all of the ways with its own data. The trojan may either replace contents in the first (odd) or second (even) group of cache sets. The spy probes both groups of cache sets, and depending on the group with higher cache access latency, the secret bits are decoded. 

\subsection{Threat Model}
\label{sec:threat}

Our attack model assumes that there is a benign victim or a malicious insider process with access to sensitive information. At the same time, there exists a spy who is trying to steal or infer the secrets. 
In this article, we demonstrate a robust attacker that does not rely on any memory sharing, and utilizes Prime+Probe-based techniques to launch attacks by creating conflict misses on cache sets. 
In the rest of this article, we use the term \emph{trojan} to refer to both victims in side channels and trojans in covert channels, since they are behaviorally similar in terms of mutual cache evictions.

\section{Cache Occupancy Changes in Timing Channels}
\label{sec:motivation} 

In this section, we describe how {\it cache occupancy patterns manifest during timing channel attacks} that provide the fundamental motivation for our framework. 

\vspace{1mm}
\noindent\textbf{Cache Occupancy Patterns for Adversaries.}
In single-group attacks, when trojan transmits a bit `1', the trojan's cache occupancy should first increase (due to trojan fetching its cache blocks) and then decrease (during spy's probe phase when trojan-owned blocks are replaced). Similarly, the spy's cache footprint would first decrease (due to trojan's filling in the cache blocks) and then increase (when spy probes and fills the cache with its own data). When trojan transmits a bit `0', neither of the processes change their respective cache occupancies. In two-group attacks, regardless of whether trojan transmits `1' or `0', we observe a gain-loss pattern in their cache occupancies.

\begin{figure}[t]
	\centering
	\includegraphics[scale=0.45]{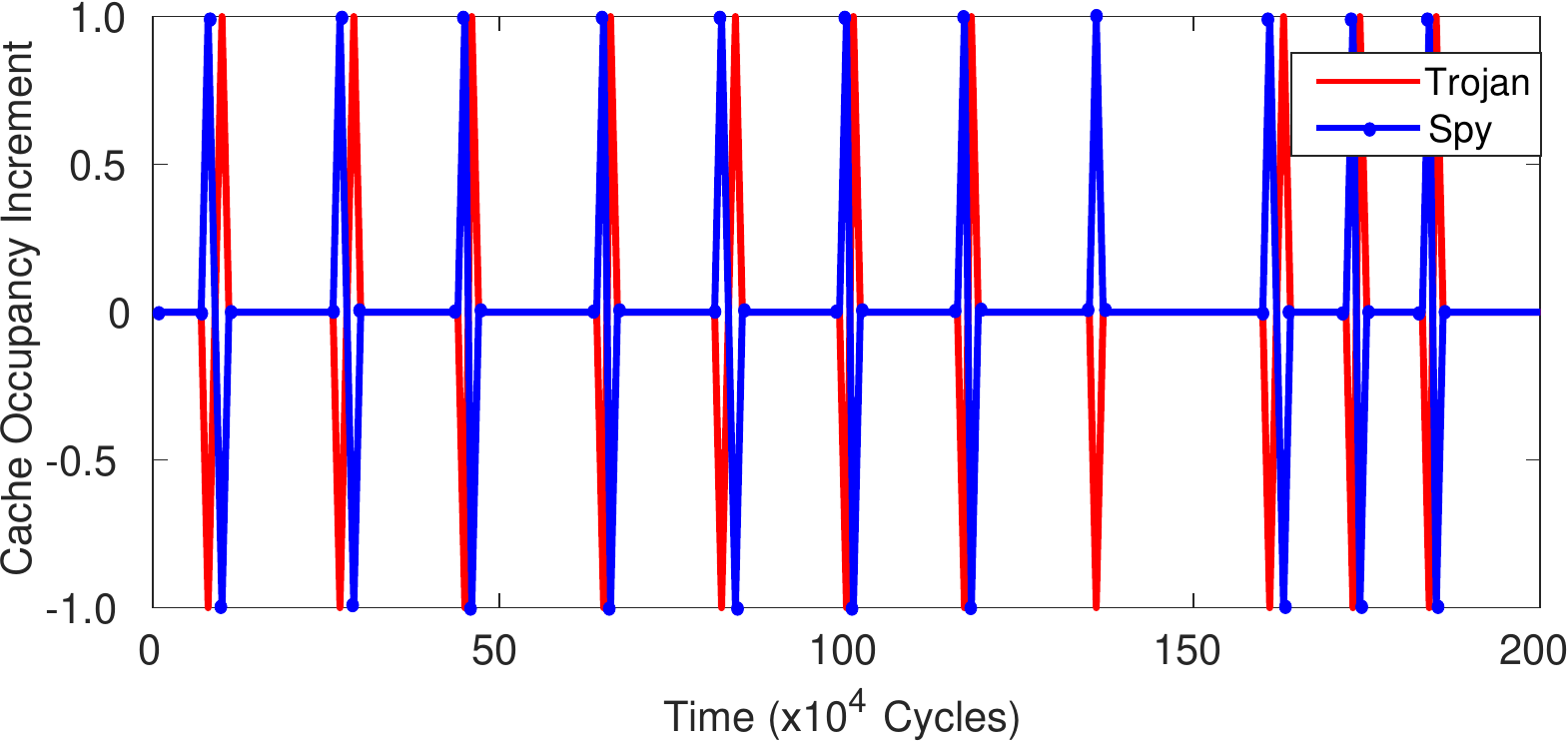}
	\vspace{-2mm}
	\caption{Cache occupancy changes for trojan and spy}
	\label{fig:motiv-attack}
	
\end{figure}

To experimentally demonstrate our observation, we implement a single group attack,
and study cache occupancy changes. 
Figure~\ref{fig:motiv-attack} shows a representative window capturing rate of change in cache occupancy over time. As we can see, the trojan's cache occupancy gain in proportion to spy's loss and vice versa. In other words, each gain in cache occupancy by the trojan is coupled by a corresponding loss of occupancy on the spy's side, and vice versa. Moreover, we can also observe \emph{a sequence repetitive pattern of gain-loss curves} in timing channels due to a series of continuous transmission.

\begin{figure}[t]
	\centering
	\includegraphics[scale=0.5]{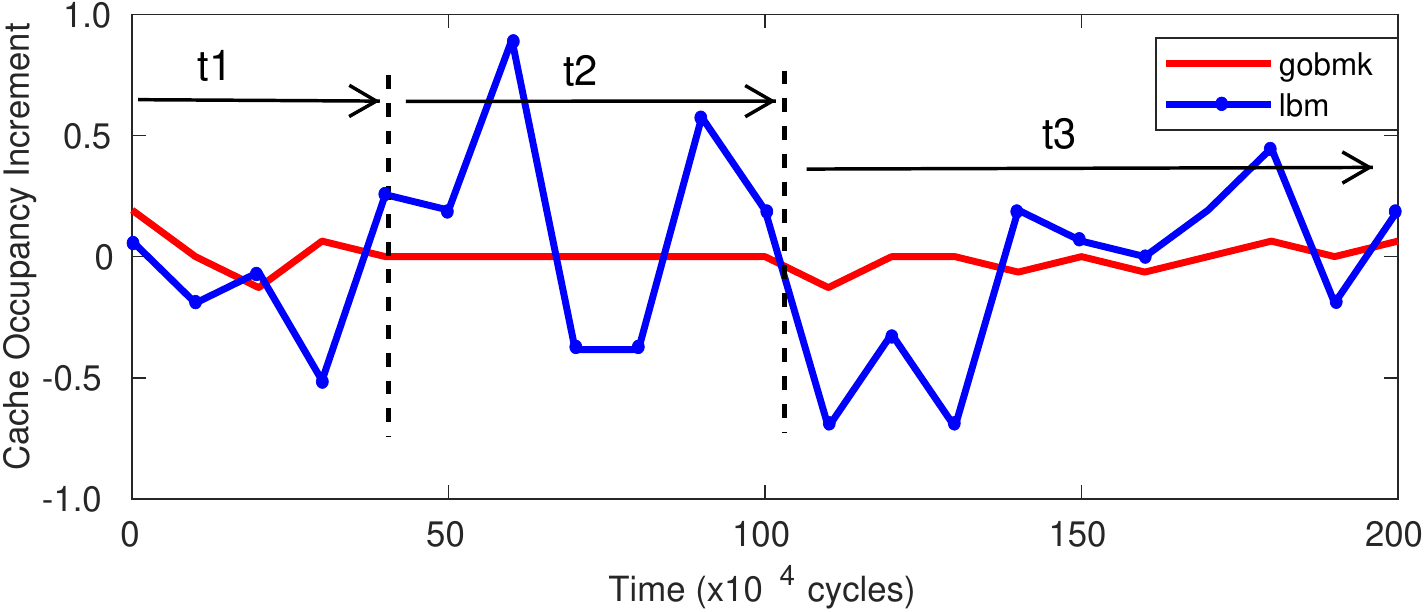}
	\vspace{-2mm}
	\caption{Cache occupancy changes for benign workloads: lbm and gobmk.}
	\label{fig:motiv-benign}
	\vspace{-4mm}
\end{figure}

\vspace{1mm}
\noindent\textbf{Cache Occupancy for Benign Workloads.}
To contrast with regular applications that have no known timing channels, we also show a representative benign application pair from SPEC2006 benchmarks~\cite{spec} with relatively high cache activity, namely \emph{lbm} and \emph{gobmk}. We observe that these application pairs do not usually show any repetitive gain-loss pattern in their occupancy traces. As we can see from Figure~\ref{fig:motiv-benign}, the occupancy patterns are rarely correlated (no obvious gain-loss patterns), e.g., there are time periods when both applications have unaligned negative dips such as in \emph{time period t1}. This may correspond to effect from a third party process's cache accesses. In \emph{time period t2}, we observe that one application's cache occupancy fluctuates while the other remains unchanged, indicating that these two benchmarks are not competing for caches mutually. Finally, there are time periods (such as in \emph{time period t3}) when the two cache occupancies almost change in the same direction. 
Overall, we can see that the cache occupancy curves between the two benign applications over time are fairly random. 

\vspace{1mm}
\noindent\textbf{Observations.}
Based on the discussion above, we make the following key observation: Timing channels in caches fundamentally rely on repetitive conflict misses that enables the trojan to influence spy's cache access timing. \emph{These conflict misses create repetitive gain-loss patterns in cache occupancy regardless of the specific timing channel implementations}. 

Unlike cache timing channel attacks, benign workloads do not intentionally create conflict misses in repetitive and controlled manner, and hence their cache occupancy patterns are sporadic and largely irregular, making them easily distinguishable from the attackers. Therefore, cache occupancy can be utilized as a useful source of information for cache timing channel detection.

\section{Why is cache occupancy a better indicator?}
\label{sec:eva}

Cache occupancy can be used as a better indicator to detect cache timing channels compared to cache miss patterns for several reasons. As discussed in Section~\ref{sec:motivation}, the cache occupancy traces for the trojan and spy pair inherently exhibit gain-loss patterns. In other words, these two patterns will have a strong negative-correlation, which can be captured with cross-correlation analysis. More importantly, cache occupancy-based analysis can use effective filtering for non-timing channel activities. Specifically, conflict misses are a fundamental mechanism to alter cache access timings. If conflicts misses happen between two processes, we could observe gains in cache occupancy for one process and simultaneous losses in occupancy for the other process (and vice versa). Other patterns such as both gains or both losses in cache occupancy are related to non-communicating activities (such as influence from third party processes), and hence can be filtered for robust cache timing channel identifications. 

In contrast, number of cache misses is an unidirectional measure within any period of time. A trojan or a spy process can intentionally issue additional memory accesses that inflate its own cache miss patterns and weaken the correlation with the other process involved. Additionally, the effect of external application activities to the cache miss traces cannot be easily separated from timing channel-related cache misses. 

\vspace{-1mm}
\section{Detecting Cache Timing Channels}
\label{sec:method}

To detect potential adversaries, we collect traces of both cache misses and cache occupancy, and compute the normalized cross-correlation (absolute value) for a pair of applications. The normalized cross-correlation has a value range of $0$ to $1$. If the value of the cross-correlation is $1$, the two traces are linearly correlated and indicate that modulation of cache accesses typically seen in trojan-spy communication. On the other hand, a close-to-zero value represents that the two traces have no correlation and the two corresponding processes are largely independent, which can be considered as lack of any explicit modulation in cache accesses. The normalized cross-correlation of two series $x$ and $y$ is:
\begin{equation}
	\gamma_{x, y}(\tau)=|\frac{1}{N}\sum_{n}\frac{1}{\sigma_x\sigma_y}(x(n)-\overline{x})(y(n-\tau)-\overline{y})|
\end{equation}
where $\sigma_x$ and $\sigma_y$ are the standard deviations of series $x$ and $y$, $\overline{x}$ and $\overline{y}$ are the means of series $x$ and $y$, and $\tau$ is the lag of the two series.

\section{Evaluation}

\begin{figure}[t]
	\centering
	\includegraphics[scale=0.28]{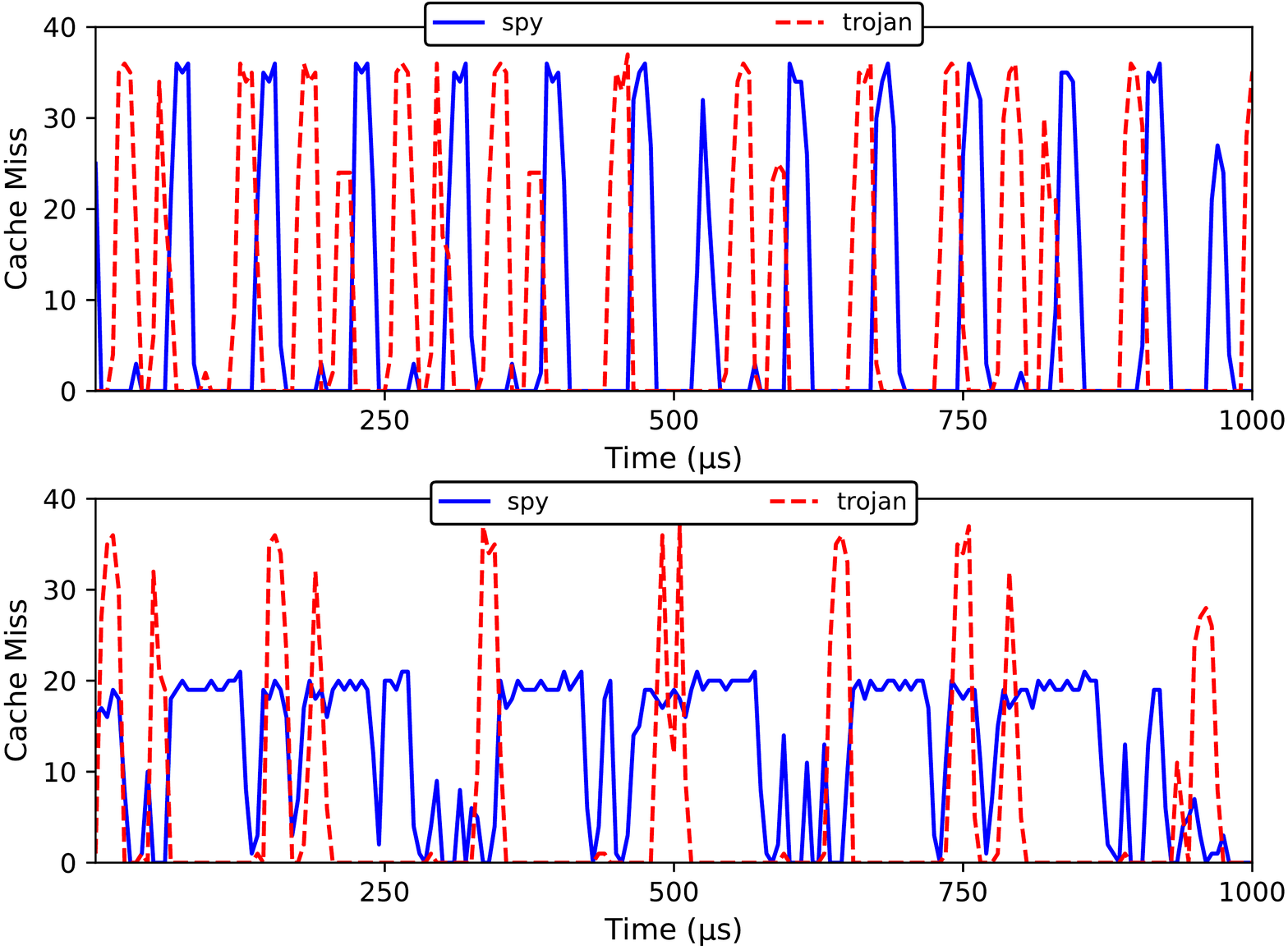}
	\vspace{-2mm}
	\caption{Cache miss patterns for trojan and spy without noise (Top) and with intentionally added noise (Bottom).}
	\label{fig:cachemiss-pattern}
	\vspace{-4mm}
\end{figure}

\subsection{Experimental Setup}
\label{sec:sim}
\noindent\textbf{Simulation Platform.} We use Gem5~\cite{gem5}, a cycle-accurate full system simulator to perform our measurements. We configure Gem5 with four x86 cores, 32KB per core private 2-way set associative L1 cache and a 512KB shared 8-way set associative L2 cache. Each application thread is pinned to a separate core using \emph{taskset}. All experiments are run in full system mode under a minimal Linux distribution with kernel version 2.6.32. We instrument the simulated cache infrastructure to support per-core cache occupancy tracking.  

\vspace{1mm}
\noindent\textbf{Runtime Correlation Analysis.} We implement a kernel thread that periodically reads per-core cache miss and occupancy values at a sampling rate of 20KHz. Note that our detection algorithm in Section~\ref{sec:method} employs a pairwise analysis in the system since the existence of trojan and spy is not known ahead of time. We note that not all pairs of application domains need to monitored, and can be limited to mutually distrusting or suspicious ones. The detection is performed every one seconds, which is significantly longer than the time to compute the cross-correlation of an application pair ($< 10ms$).

\vspace{1mm}
\noindent\textbf{Cache Timing Channel Attacks}. We implement a prime+probe covert channel attack on last level cache (L2 cache) similar to those prior works~\cite{whispers,liu2015last}. 
We generate 32 sets of addresses for the trojan and spy. Addresses in the same set will map to the same cache set, which would be used to create conflict misses. We configure the trojan and spy to utilize a single group of cache sets. Note multi-group attacks will exhibit similar gain-loss cache occupancy patterns (as discussed in Section~\ref{sec:motivation}).

\begin{figure}[t]
	\centering
	\includegraphics[scale=0.28]{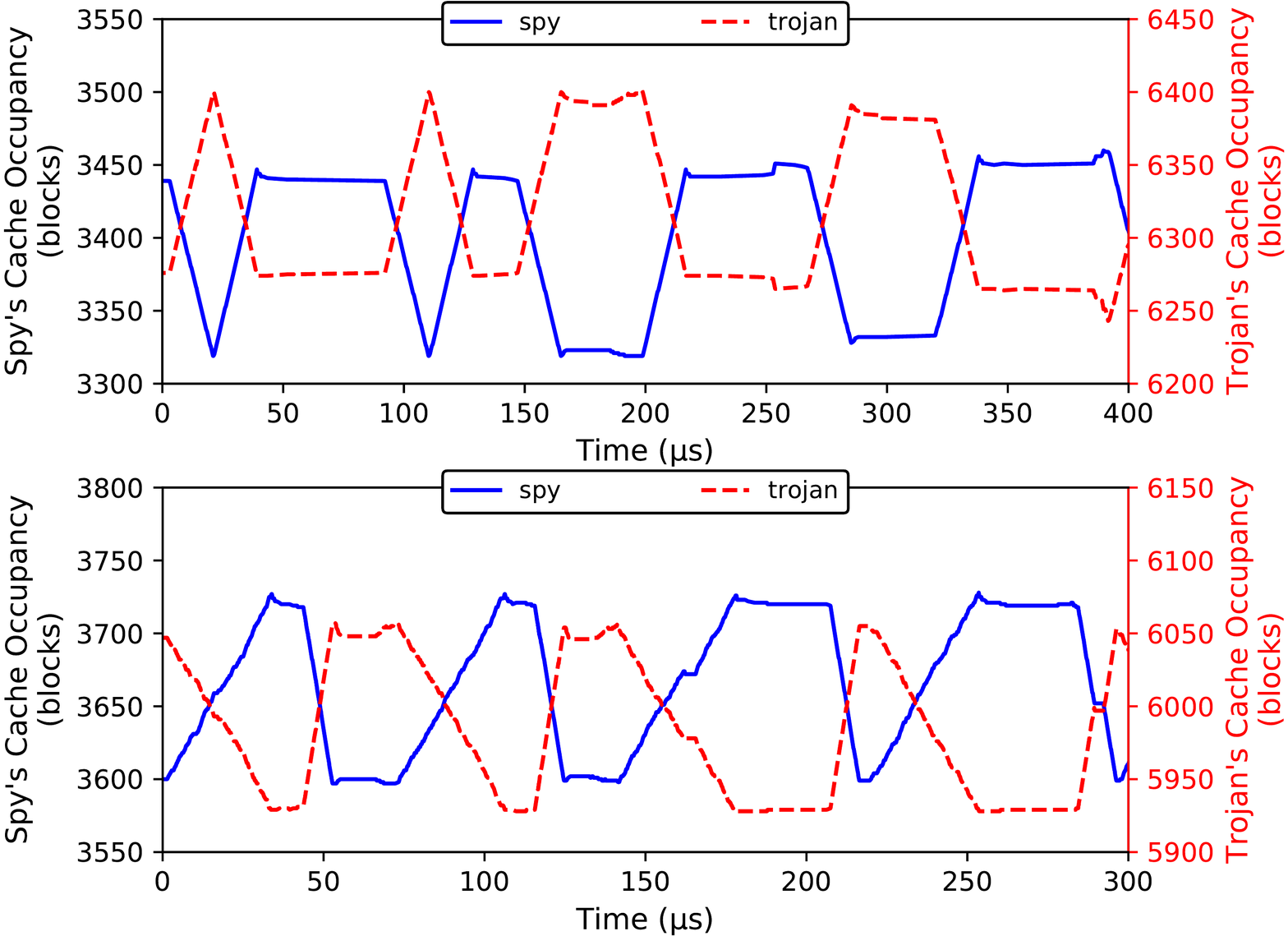}
	\vspace{-2mm}
	\caption{Cache occupancy patterns for trojan and spy without noise (Top) and with intentionally added noise (Bottom).}
	\label{fig:occupancy-pattern}
	\vspace{-4mm}
\end{figure}
\subsection{Identifying Sophisticated Adversaries}
\label{sec:results}

We implement a naive and a sophisticated trojan-spy pair with the following setup: 

\begin{itemize}
	\item The two processes dynamically generate certain sets of conflict addresses (that map to 32 cache sets).
	\item The selected cache sets are split into two parts. The first 16 sets are used in timing channel communication and the second half will be utilized for random noise injection.
	\item In a naive case, the trojan and spy merely exploit the 16 sets earmarked for covert communication. For smart adversaries, during each iteration of the spy's prime and probe phase, it simultaneously accesses a random number of sets with random intervals to create additional cache misses. 
\end{itemize}

\begin{figure}[t]
	\centering
	\includegraphics[scale=0.28]{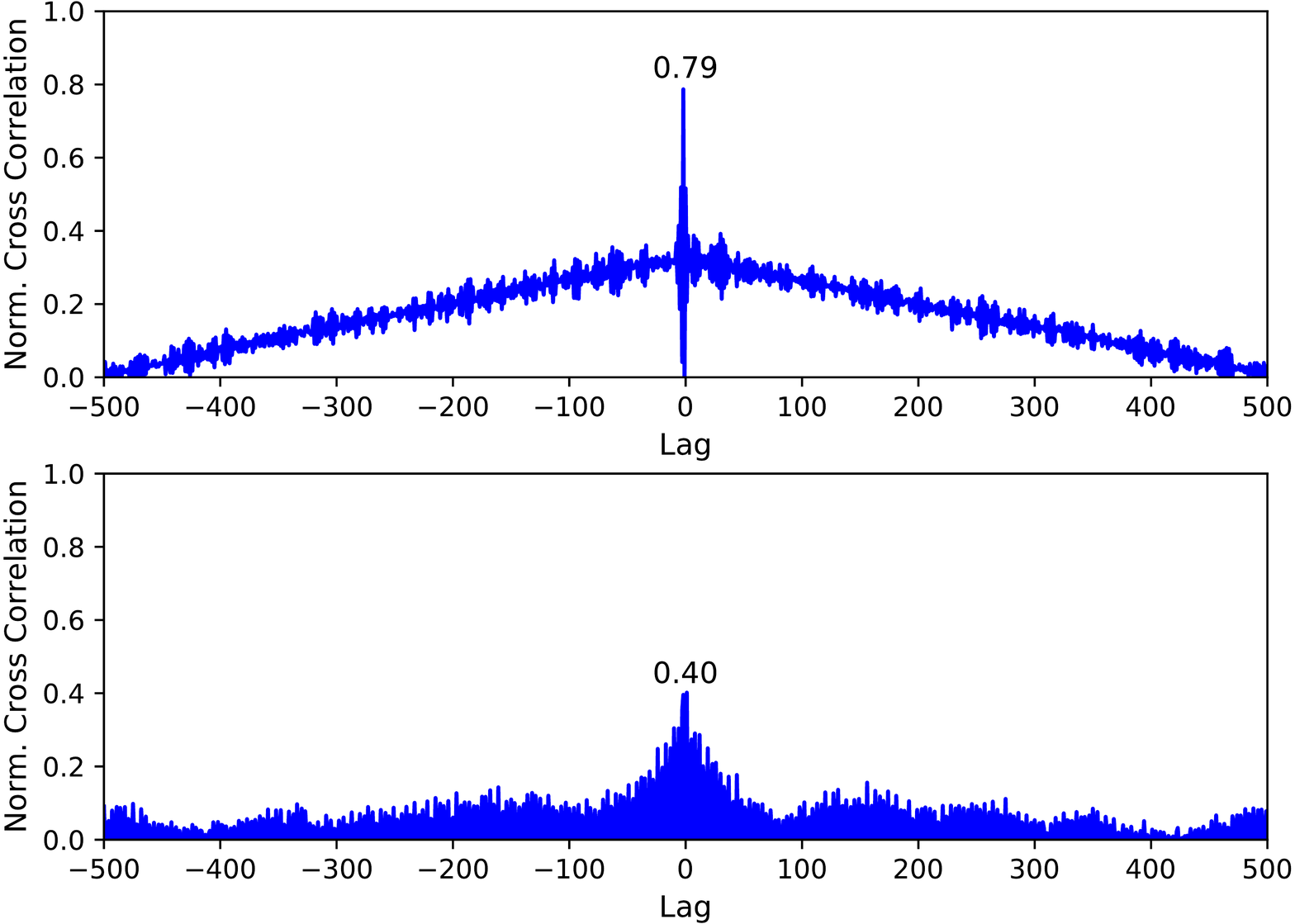}
	\vspace{-2mm}
	\caption{Cross-correlation between miss patterns with naive, noise-free trojan-spy (Top) and sophisticated trojan-spy with injected noise patterns in cache misses (Bottom).}
	\label{fig:cachemiss-cc}
	\vspace{-4mm}
\end{figure}

\begin{figure}[t]
	\centering
	\includegraphics[scale=0.28]{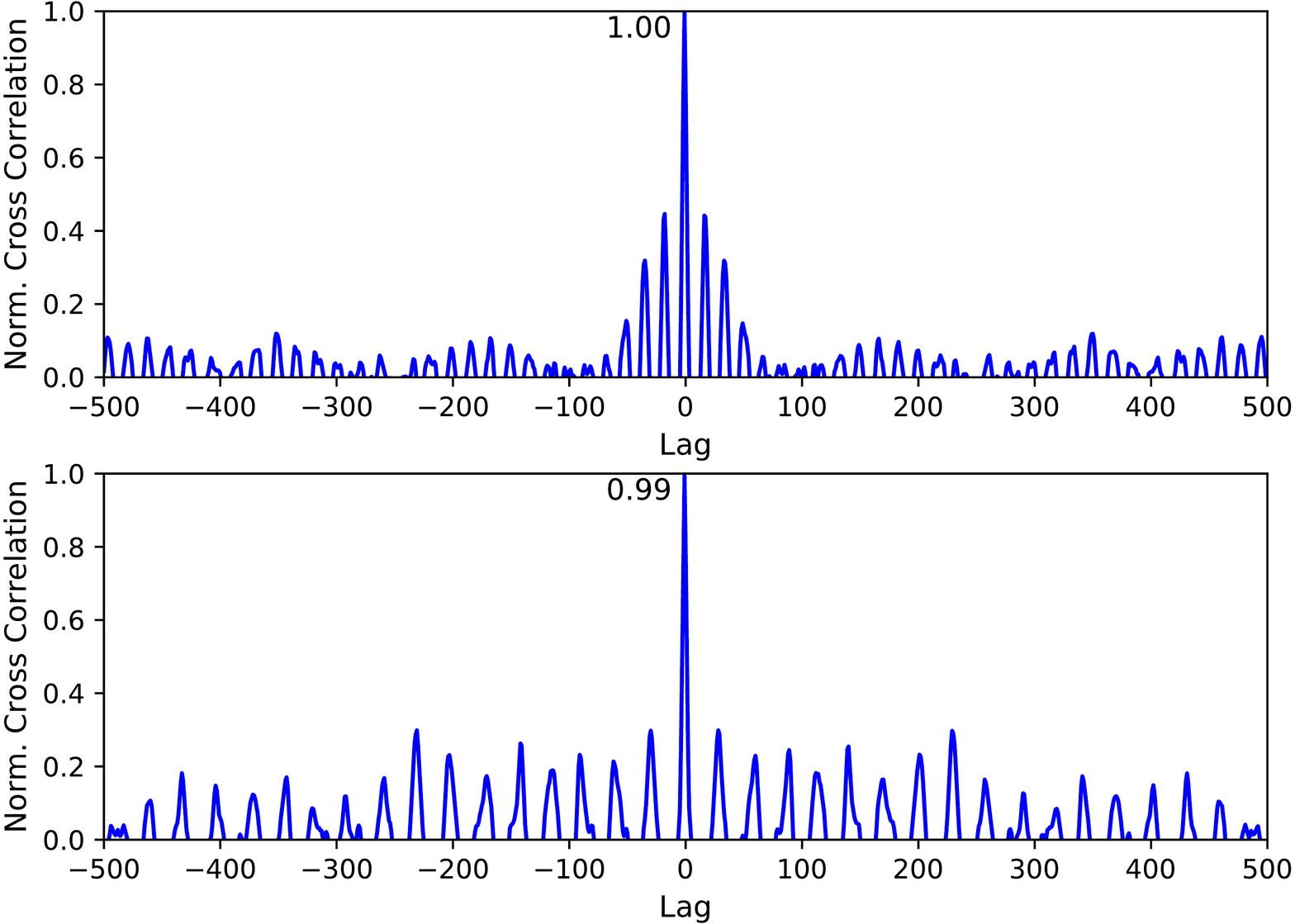}
	\vspace{-2mm}
	\caption{Cross-correlation between occupancy patterns with naive, noise-free trojan-spy (Top) and sophisticated trojan-spy with injected noise patterns in cache misses (Bottom).}
	\label{fig:occupancy-cc}
	\vspace{-4mm}
\end{figure}

We sample the L2 cache misses as well as the cache occupancy with a sampling window of 10,000 cycles. Figure~\ref{fig:cachemiss-pattern} shows the cache miss patterns for the naive and sophisticated adversaries. 
When no intentional noise is injected, we observe multiple peaks in cache misses interleaved between the trojan and spy (top plot in Figure~\ref{fig:cachemiss-pattern}). 
In most cases, a sharp peak from the trojan is followed by a very similar sharp peak from the spy, which corresponds to trojan's eviction in the prime/probe and spy's eviction when it accesses the same cache sets. However, after the noise is injected, the strongly coupled cache miss peaks from the trojan and spy no longer exist. We observe that the spy has cache misses that differ in both time and space dimension. The continuous cache miss pattern in the spy makes it almost non-correlated with the trojan (bottom plot in Figure~\ref{fig:cachemiss-pattern}). Figure~\ref{fig:occupancy-pattern} demonstrates the cache occupancy traces for the same attacks. In the naive scenario, we can see the strong gain-loss pattern shown as the mirroring cache occupancy from the two processes. Even with the smart adversaries, we are still able to observe similar patterns due to the fact that cache occupancy only records mutual behavior in cache block gain and losses. More importantly, it is immune to noise effects from self-eviction and interference from third party processes (that will reflect as mutual cache occupancy change between that process and the adversary).
We compute the cross-correlation between the trojan and spy using both the cache miss and the cache occupancy patterns (See Section~\ref{sec:method}). Figure~\ref{fig:cachemiss-cc} and Figure~\ref{fig:occupancy-cc} illustrate the two sets of normalized cross-correlation values at various lags. The value shown in each of the plots represents the highest correlation value observed.
We can see that using cache miss patterns, there is a relative high correlation (\textbf{0.79}) between trojan and spy in the naive attack. However, when intentional noise is injected by the spy in the form of additional cache misses, the value drops significantly to \textbf{0.4}, which successfully obfuscates cross-correlation and thus hides the traces of communications. On the other hand, cache occupancy-based analysis exhibits much higher robustness in indicating the presence of cache timing channels: First, in the naive attack, the correlation in trojan-spy occupancies is even higher compared to that in cache misses (\textbf{1.0}); More importantly, there is very minimal change ($<$\textbf{0.01}) in occupancy-based correlation after additional cache misses are injected intentionally by sophisticated adversaries. The results clearly show that cache occupancy can be leveraged as a stronger  evidence for detecting cache timing channels. 

Interestingly, we also find that the cross-correlation peaks at a certain non-zero lag for cache miss traces (in Figure~\ref{fig:cachemiss-cc}), but always peaks with zero lag for cache occupancy traces. This is because there is a delay in time corresponding to trojan's and spy's cache miss events during timing channels. However, trojan and spy's occupancy traces change simultaneously (temporally, spy's loss is trojan's gain and vice versa). Hence, they always align as we can see from Figure~\ref{fig:cachemiss-pattern} and Figure~\ref{fig:occupancy-pattern}. This indicates the potential benefits of occupancy-based analysis in terms of speed as only 0-lag has to be calculated to determine the maximum cross-correlation.

\subsection{Discussion}
\label{sec:discussion}

A potential way for an individual attacker (either trojan or spy) to fuzz the suspicious cache occupancy pattern is through explicitly issuing the \emph{clflush} instruction that decreases its cache occupancy by itself. Note that normal load and store instructions can only increase the cache occupancy. In this case, \emph{clflush} instructions can be audited to obtain the cache occupancy changes associated with the existence of timing channels. Also, in some systems, \emph{clflush} instruction is disabled from user-space due to various security concerns (such as Google's Native Client Sandbox~\cite{google-rowhammer}). 

\section{Related Work}

Cache side- and covert timing channels have been demonstrated on real hardware in prior studies~\cite{ristenpart2009hey,whispers,yarom2014flush+,xu2011exploration,liu2015last,yao2017covert,yao_hpca18,alagappan2017dfs,yao2018covert}.  

Venkataramani et al.~\cite{guru2016} propose a generic framework that detects cache covert timing channel using correlations between cache conflict misses. ReplayConfusion~\cite{replayconfusion} records and replays cache access traces from the trojan and spy, and detects cache timing attacks based on differences of cache misses under different cache slice hashing functions. Fang et al~\cite{fang2018prefetch} have demonstrated novel use of hardware prefetchers to stop cache timing channels. These techniques require customized hardware for either fine-grained conflict miss tracking or cache design changes. 

Additionally, Demme et al. propose a machine learning-based approach that identifies cache side channels based on performance counters~\cite{demme_2013}. We note that as a stronger indicator, cache occupancy observations can be leveraged together for more robust detection. Meanwhile, our detection framework using cache occupancy can be potentially incorporated with existing prevention techniques such as cache partitioning~\cite{wang2016secdcp,kirianskydawg} to build end-to-end cache timing channel defense solutions. In addition to improving cache resiliency, we note that other mechanisms to improve software robustness~\cite{simber,sarre,tradeoffs,deft,lime,statsym} and memory reliability~\cite{chen2012repram,chen_jetc14} will improve overall system security. 

\section{Conclusion}
\label{sec:concl}
In this work, we motivate that cache occupancy traces can be utilized as a better indicator for cache timing channel detection. We demonstrated that cache access latency modulation between a trojan and spy generates  gain-loss cache occupancy patterns. 

Our results have shown the trojan and spy's malicious activities can be successfully recognized even when the adversaries intentionally inject noise (that effectively hide its traces in cache miss patterns). Our works highlighted that the new cache event can be used to build robust and effective cache timing channel detector in the future.

\bibliographystyle{plain}
\bibliography{refs} 
\end{document}